\documentclass{article}
\usepackage{arxiv}
\usepackage[utf8]{inputenc} 
\usepackage[T1]{fontenc}    
\usepackage{hyperref}       
\usepackage{url}            
\usepackage{dsfont}
\usepackage{booktabs}       
\usepackage{multirow}
\usepackage{amsfonts}       
\usepackage{nicefrac}       
\usepackage{microtype}      
\usepackage{lipsum}		
\usepackage{graphicx}

\usepackage[numbers]{natbib}
\usepackage{doi}
\usepackage{caption}
\usepackage{amsmath,amsfonts,bm}

\newcommand{\rr}{\mathbb{R}}

\newcommand{\nn}{\mathbb{N}}

\title{Accelerated Hydration Site Localization and Thermodynamic Profiling}

\author{
Florian B. Hinz
\thanks{Department of Pharmaceutical Sciences, University of Basel, Basel, Switzerland}
\ \thanks{Swiss Institute of Bioinformatics, Basel, Switzerland}
\ \thanks{These authors contributed equally to this work.}
\And
Matthew R. Masters \footnotemark[1] \ \footnotemark[2] \ \footnotemark[3] \And
Julia N. Kieu \footnotemark[1] \And
Amr H. Mahmoud \footnotemark[1] \ \footnotemark[2] \And
Markus A. Lill \footnotemark[1] \ \footnotemark[2] \thanks{Corresponding Author: markus.lill@unibas.ch} \And
}

\begin{document}

\maketitle

\begin{abstract}
Water plays a fundamental role in the structure and function of proteins and other biomolecules.
The thermodynamic profile of water molecules surrounding a protein are critical for ligand binding and recognition.
Therefore, identifying the location and thermodynamic behavior of relevant water molecules is important for generating and optimizing lead compounds for affinity and selectivity to a given target.
Computational methods have been developed to identify these hydration sites, but are largely limited to simplified models that fail to capture multi-body interactions, or dynamics-based methods that rely on extensive sampling.
Here we present a method for fast and accurate localization and thermodynamic profiling of hydration sites for protein structures.
The method is based on a geometric deep neural network trained on a large, novel dataset of explicit water molecular dynamics simulations.
We confirm the accuracy and robustness of our model on experimental data and demonstrate it's utility on several case studies.
\end{abstract}

\section{Introduction}
Water plays a fundamental role in the structure and function of proteins and other biomolecules.
The ubiquitous presence of water in protein environments significantly impacts the stability, dynamics, and interactions with other molecules.
In particular, the desolvation of hydrophobic moieties is known to be a major driving force for ligand binding and recognition.
Water molecules may also mediate interactions between ligand and protein, through the formation of water bridges.
Therefore, a complete understanding of the effect of solvation is essential to optimize the binding affinity and selectivity of lead compounds.
Beyond small molecules, water also plays an important role in mediating other interactions such as protein-protein and other biomolecular interactions, which are becoming increasingly important with the development of new biologic products.

Due to it's importance and range of potential applications, numerous computational methods have been developed in order to predict the location of hydration sites, localized regions of high water density.
However, these methods are usually limited to simplified models that fail to capture the dynamic behavior and multi-body interactions inherent in water networks, or dynamics-based methods that rely on extensive sampling that is computationally prohibitive for flexible proteins with significant conformational change.
Furthermore, simply identifying the position of likely water molecules is insufficient for ligand optimization since the thermodynamic profile of individual water molecules dictates whether displacement by a ligand is energetically favorable or not.
Computational methods also exist for estimating the thermodynamic properties of hydration sites but suffer from the same accuracy/cost trade-off between static and dynamic-based methods.

In this paper, we introduce a generative deep learning model that is capable of fast and accurate hydration site identification and thermodynamic profiling.
Unlike previous methods, our model is capable of predicting hydration sites in one-shot with near dynamics-level accuracy.
The model is able to resolve complex multi-body interactions in a fixed-time and propose physically valid water networks along with associated enthalpic and entropic contributions.
It is based on an equivariant transformer network and is trained to match the explicit distribution of hydration sites on a protein surface.
Training was done using a novel dataset of WATsite simulation results from thousands of diverse protein systems, representing a high level of robustness and generalizability across protein space.
Our method was validated using a difficult hold-out set of unseen protein sequences in addition to high-quality crystallographic waters.
We further demonstrate and discuss our model's utility on a number of interesting drug and protein design scenarios.

\section{Background and Related Methods}
\subsection{Experimental Determination}
Water locations can be resolved using several experimental methods including x-ray/neutron crystallography \cite{savage198611}, cryo-EM \cite{halle2004protein, henderson1975three}, and NMR \cite{wuthrich2001way}.
Crystallography is the most common technique with more than half of the experimental structures containing at least one water molecule \cite{zsido2023advances}.
However, there are a number of technical challenges associated with the determination of water locations leading to certain limitations.
Most of these challenges arise from the complex multi-body interactions and high mobility inherent to water molecules.
Experimental determination is often limited to the first shell of waters, especially on the solvent-exposed surface and not buried within the protein.
Furthermore, experimental approaches have limited resolution and water positions often have non-ideal interaction geometry.
Lastly, experimental determination of water locations does not quantify thermodynamics properties such as enthalpy and entropy.

\subsection{Computational Methods}
While experimental determination of water structure is indispensable for providing the ground truth, they are limited by real-world constraints such as time, cost, and technical feasibility.
To overcome these challenges, a number of computational methods have been developed to predict the location of water molecules around an existing protein structure.
These methods can be largely grouped into two categories, static and dynamic.
Static methods assume the water network has a constant position and neglects any slowly-resolving water-water or water-protein interactions as well as any protein movement or flexibility.
Among the static methods, most fall into the category of knowledge-based, such as 3D-RISM \cite{kovalenko1998three,kovalenko1999self,nittinger2019water}, AQUARIUS \cite{pitt1991modelling}, SZMAP \cite{bayden2015evaluating}, and WaterFLAP \cite{nittinger2019water, mason2013high, baroni2007common}, geometry-based, such as AcquaAlta \cite{rossato2011acquaalta}, Auto-SOL \cite{vedani1991algorithm}, HADDOCK \cite{van2006solvated}, WarPP \cite{nittinger2018placement}, and WATGEN \cite{bui2007watgen}, or energy-based, such as Fold-X \cite{schymkowitz2005prediction} or WaterDock \cite{sridhar2017waterdock}. 

Dynamic methods often employ molecular dynamics (MD) simulations which utilize an explicit water model and protein force field.
By using MD simulations, one can generate a trajectory detailing the atomic behavior of water with other molecules in solution.
This trajectory can be later analyzed to identify regions of high water density, characterize the thermodynamics of these sites, and understand detailed interactions involving water.
However, the limitation of dynamic methods is that they are computationally and time intensive, especially due to the calculation of large number of non-bonded interactions that is required by the explicit water simulation.
Furthermore, in order to ensure convergence of the water density within a reasonable time, strong positional constraints are added to the protein solute, removing any flexibility and perturbation to the initial protein structure \cite{hu2012protein, yang2014analysis}.
While these constrained results can still be useful in guiding structure-based drug design for particular targets, we know that protein flexibility plays a significant role in most drug targets and that even small perturbations to protein structure can lead to drastically different hydration shell structure \cite{yang2016dissecting}.
Therefore, dynamic methods are still largely limited to single protein structures and require either multiple runs or time-consuming, extensive runs in order to achieve proper convergence.
Among the dynamic methods are HydraMap \cite{li2020prediction}, HyPred \cite{virtanen2010modeling}, MobyWat \cite{jeszenoi2016exploration,jeszenHoi2015mobility}, WaterMap \cite{abel2008role,nittinger2019water}, and WATsite \cite{hu2014watsite,yang2017watsite2,yangwatsite}.

More recently, some works have employed machine learning and deep neural networks towards the prediction of hydration shell structure.
For example, several groups have used convolutional neural networks that work on 3D grids in order to predict the water density such as GalaxyWater-CNN \cite{park2022galaxywater}, HydraProt \cite{zamanos2024hydraprot}, and Ghanbarpour et al., 2020 \cite{ghanbarpour2020instantaneous}.
In another work, Park developed a SE(3)-equivariant network that is capable of directly predicting positions of water molecules given pre-defined probe locations \cite{park2024water}.
Lastly, SuperWater is a diffusion-based model designed to generate crystal waters given the protein structure \cite{superwater2024}.
While these methods take a step towards the development of a one-shot hydration site predictor, they still fall short by either training on sparse and poorly modeled crystallographic data or by not considering the thermodynamic contributions of hydration sites.

\section{Methods}
The prediction of the hydration sites with associated thermodynamic properties consists of two separate models. The first model predicts the hydration site location and is explained in section \ref{sec:location_prediction}. The predicted coordinates then serve as an input to the second model, which predicts the associated entropy and enthalpy values (see section \ref{sec:enthalpy_entropy_prediction}). The code and a reference to the training data are provided at \href{https://github.com/lillgroup/HydrationSitePrediction}{https://github.com/lillgroup/HydrationSitePrediction}.

\subsection{Location prediction}
\label{sec:location_prediction}

\begin{figure}[ht]
\centering
\includegraphics[width=0.8\linewidth]{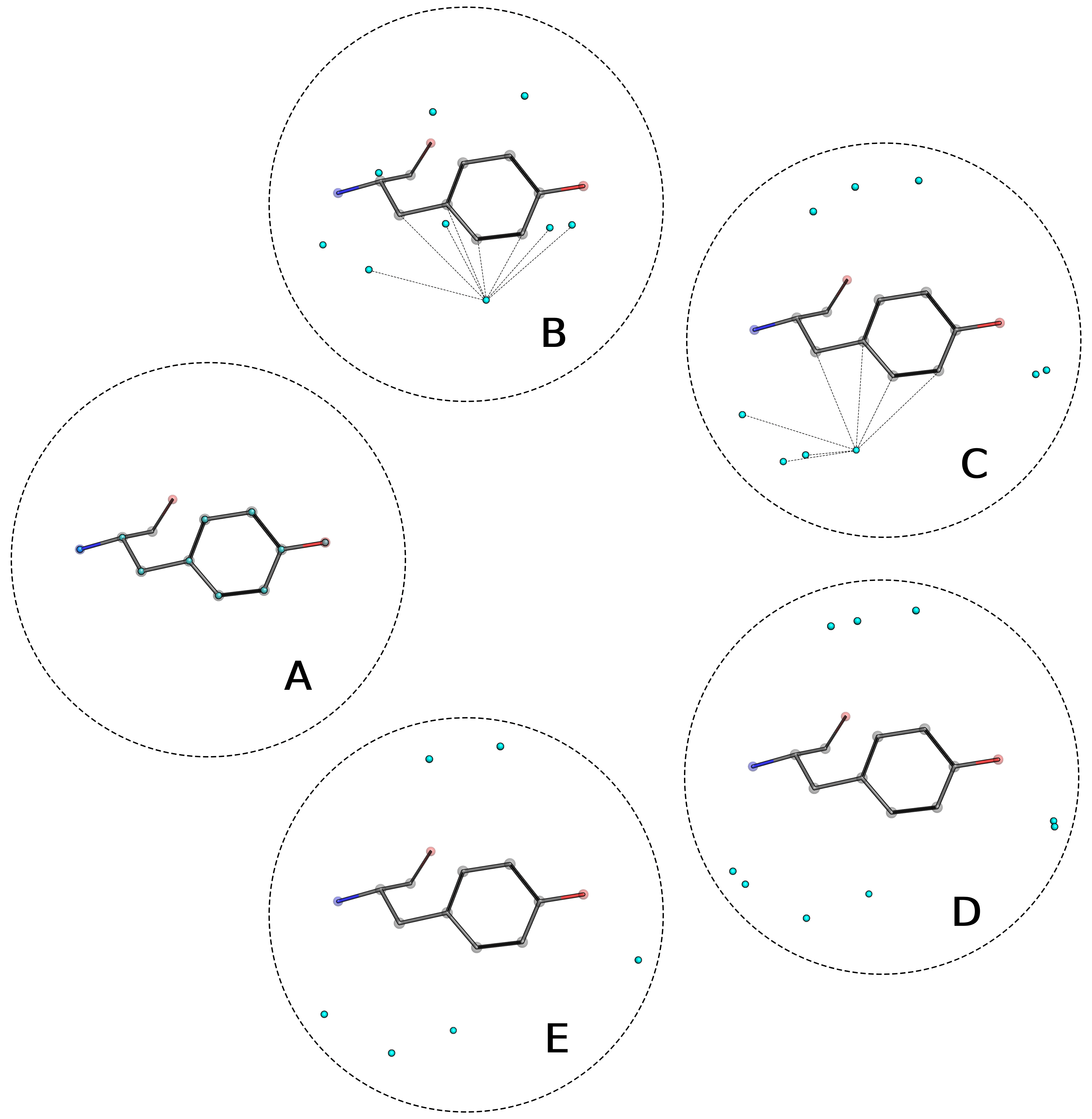}
\caption{Schematics displaying the coordinates prediction of hydration sites. \textbf{A}: Initially, we position a water prediction (light blue spheres) at every atom exceeding a certain cutoff of solvent exposure (SASA greater than $0.1$). An equivariant attention layer is applied to the distance-based graph topology to perturb the water predictions. \textbf{B, C}: The distance based graph topology is updated by now also considering water predictions as graph nodes. An equivariant attention layer perurbs the water predictions. \textbf{D}: After the application of five layers, predictions of low certainty are filtered out and a clustering algorithm is applied to end up at the final water predictions (\textbf{E}).}
    \label{fig:location_prediction}
\end{figure}

The procedure of learning the hydration site locations is inspired by our previous work \cite{Hinz_2023} and schematically shown in Figure \ref{fig:location_prediction}. The input data consists of the atom coordinates and the associated feature vectors. The feature vector is $92$ dimensional consisting of a one-hot encoding of the atom type ($38$ dimensions), a one-hot encoding of the residue type ($21$ dimensions), a radial basis function expansion of the SASA value ($32$ dimensions) and a one-hot encoding for the node type (atom or predicted water, $1$ dimension). Initially, $n \in \nn$ water predictions are placed at the position of any atom of a SASA value greater than $0.1$ (see Figure \ref{fig:location_prediction}A). A graph structure is constructed from the atom positions by connecting any two nodes (i.e. atoms) via an edge, if their distance is less than $6$\AA. The edge feature encoding consists of a radial basis function expansion of the edge length ($32$ dimensions). We apply a first equivariant attention layer to the graph structure to obtain updated locations of the water molecule nodes. The protein atom positions are left unchanged. The graph topology is updated by also including the nodes representing the water molecules and again constructing edges between nodes less than $6$ A apart. Five equivariant attention layers are applied to perturb the nodes representing water molecules, with the distance-based graph topology being updated before applying each layer (see B and C in Figure \ref{fig:location_prediction}). The network is trained by minimizing a loss function that resembles the Kullback-Leibler divergence between two gaussian mixture models: Once the mixture components are defined via the ground truth water coordinates and once via the predicted coordinates. Details about the loss function are provided in the following paragraph.

\subsubsection{Loss function}
The equivariant graph neural network as discussed in section \ref{sec:location_prediction} outputs tuples $(x_{j},w_{j})\in \rr^{3}\times [0,1] $ for $j\in \left\{ 1,\cdots,n \right\} $ and $n\in \nn$ the number of initially placed water prediction nodes.
The first element $x_{j}\in \rr^{3}$ represents the coordinates of the predicted hydration site (see D in Figure \ref{fig:location_prediction}), while $w_{j}\in [0,1]$ corresponds to an associated certainty weight of the prediction. Defining for $j\in \left\{ 1,\dots,n \right\}$ the normalized certainty weights as
\begin{equation}
	q_{j}:=\frac{w_{j}}{\sum_{i=1}^{n} w_{i}},
	\label{eqn:normalized_weights}
\end{equation}
we consider the tuples $\left((x_{j},q_{j})\right)_{j=1}^{n}$ as a representation of components of a Gaussian mixture distribution as follows:
\begin{equation}
	q(x):= \sum_{j=1}^{n} q_{j} \frac{1}{\sqrt{(2\pi)^{3}} \sigma^{3}} \exp\left(-\frac{1}{2 \sigma^{2}} ||x-x_{j}||_{2}^{2} \right),
	\label{eqn:gaussian_mixture_predicted}
\end{equation}
where the standard deviation is chosen as $\sigma=0.5$. 
For a given protein in the training set, let us denote the true $m \in \nn$ hydration sites with associated occupancies of at least $0.5$ as tuples $(u_{j},o_{j})$ for $j\in \left\{ 1,\dots,m \right\}$. Similarly as in \ref{eqn:normalized_weights} and \ref{eqn:gaussian_mixture_predicted} we define
\begin{equation}
	p_{j}:=\frac{o_{j}}{\sum_{i=1}^{n} o_{i}},
	\label{eqn:normalized_weights_occupancy}
\end{equation}
and consider the true hydration sites representing a Gaussian mixture distribution as follows
\begin{equation}
	p(x):= \sum_{j=1}^{m} p_{j} \frac{1}{\sqrt{(2\pi)^{3}} \sigma^{3}} \exp\left(-\frac{1}{2 \sigma^{2}} ||x-u_{j}||_{2}^{2} \right).
	\label{eqn:gaussian_mixture_ground_truth}
\end{equation}

As a simplified surrogate for the symmetrized Kullback-Leibler divergence $\text{KL}(p|\,q)+\text{KL}(q|\,p)$, we choose 
\begin{equation}
	L_{1}\left(\left( (x_{j},q_{j}) \right)_{j=1}^{n},\left((u_{j},p_{j} )\right)_{j=1}^{m}  \right):=\sum_{j=1}^{m} p_{j} \log\left(\frac{p_{j}}{q(u_{j})} \right)+\sum_{j=1}^{n} q_{j} \log\left(\frac{q_{j}}{p(x_{j})} \right).
\end{equation}
We found it beneficial to introduce a further loss term, penalizing the concentration of weights:

\begin{equation}
	L_{2}((q)_{j=1}^{n}):= \sum_{j=1}^{m}q_{j}^{2}.
\end{equation}
The total loss function consists of a weighted sum of $L_{1}$ and $L_{2}$:
\begin{equation*}
	L_{3}\left(\left( (x_{j},q_{j})_{j=1}^{n} \right),\left((u_{j},p_{j} )_{j=1}^{m} \right) \right)=L_{1}\left(\left( (x_{j},q_{j}) \right)_{j=1}^{n},\left((u_{j},p_{j} )\right)_{j=1}^{m}  \right)+\alpha \cdot L_{2}\left((q)_{j=1}^{n}\right),
\end{equation*}
for some hyperparameter $\alpha>0$.
\subsubsection{Clustering}
At inference time, we first filter out predictions with certainty weights lower than a cutoff $w_{c}=0.035$. To the remaining predictions we apply an agglomerative clustering as implemented in \cite{scikit-learn} with a linkage distance threshold of 2\AA\ to end up at well separated predictions (see E in Figure \ref{fig:location_prediction}). Let
$M,k, k_{1},\dots,k_{M} \in \nn$ and 
\begin{equation*}
	\mathcal{C}_{k}=\left\{ (x_{k_{1}},w_{k_{1}}),\dots,(x_{k_{1}},w_{k_{M}}) \right\},
\end{equation*}
be the k'th cluster determined by the algorithm. We define the weighted cluster mean with associated certainty weight as
\begin{equation*}
(x_{\mathcal{C}_{k}}, w_{C_{k}})=\left( \sum_{j=1}^{M} w_{k_{j}} x_{k_{j}},\sum_{j=1}^{M} w_{k_{j}}\right).
\end{equation*}

We consider $x_{C_{k}}\in \rr^{3}$ a predicted hydration site location (see \ref{fig:location_prediction}D) if the associated certainty weight $w_{C_{k}}$ exceeds a threshold of $0.1$.

\subsection{Entropy and enthalpy prediction}
\label{sec:enthalpy_entropy_prediction}

\begin{figure}[ht]
    \centering
	\includegraphics[width=\linewidth]{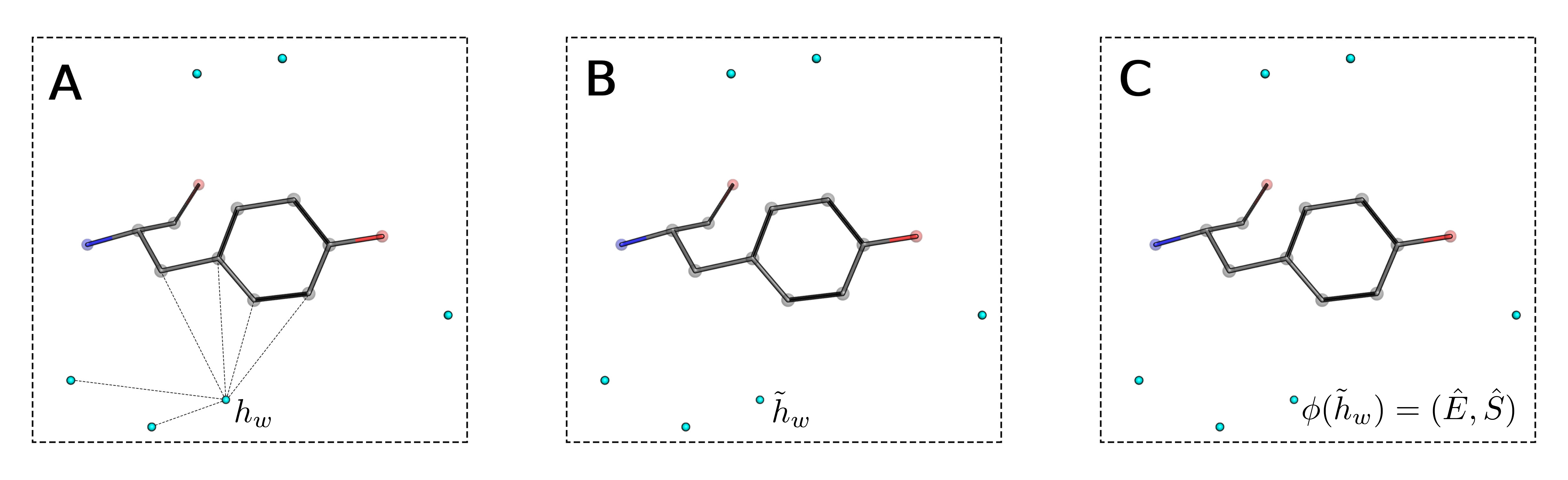}
    \caption{Schematics displaying the prediction of entropy and enthalpy. \textbf{A}: The input consists of coordinates and feature vectors for protein atoms and hydration sites. The edges are constructed based on distances. \textbf{B}: A graph attention network is applied and outputs a modified feature vector per graph node. \textbf{C}: A feed forward network is applied to the feature vectors of the water nodes. The output consists of the enthalpy and entropy values.}
    \label{fig:thermo_prediction}
\end{figure}

Figure \ref{fig:thermo_prediction} schematically shows the learning of the entropy and enthalpy values of each hydration site.  First, a graph is constructed with nodes representing atoms and hydration sites. During training, true hydration sites from WATsite with occupancy values greater than 0.3 are considered nodes. The edges are constructed based on a distance cutoff of 8\AA, ignoring intramolecular edges within the protein. The feature vectors associated with the nodes are similarly constructed as for the location prediction model (see section \ref{sec:location_prediction}). The feature vectors are updated by applying three layers of a graph attention network as introduced in \cite{gat} (see Figure \ref{fig:thermo_prediction} B). A feed forward network is applied to each feature vector to end up at a $2$ dimensional output representing entropy and enthalpy (see Figure \ref{fig:thermo_prediction} C).
The loss function during training consists of the sum of the mean squared errors of entropy and enthalpy over all hydration sites with occupancy of at least $0.5$:

\begin{equation*}
L_{\text{thermo}}\left((\hat{E}_{i},E_{i})_{i=1}^{n}, (\hat{S}_{i},S_{i})_{i=1}^{n} \right)	= \frac{1}{n}\sum_{i}^{n}\left( E_{i}-\hat{E}_{i} \right)^{2}+\left( S_{i}-\hat{S}_{i} \right)^{2},
\end{equation*}
where $n \in \nn$ is the number of hydration sites with occupancy of at least $0.5$ and
$\hat{E}_{i}$ denotes the model prediction for the target value enthalpy $E_{i}$ for the hydration site $i\in \left\{ 1,\dots,n \right\}$. Similarly, the prediction of the target value entropy $S_{i}$ is denoted by $\hat{S}_{i}$ for $\left\{ 1,\dots,n \right\}$.

\subsection{Dataset}
The training data was prepared using WATsite \cite{hu2014watsite, masters2018efficient} and the number of proteins contained in the dataset was 4148, with occupancy and thermodynamic values for the HS. For the selection of the dataset, 19443 proteins from PDBbind \cite{wang2005pdbbind, su2018comparative} were grouped by their UniProt ID and one representative structure was randomly picked from the group to prevent redundancy. The test set was selected to be as different from the validation and train set as possible, to evaluate the model’s ability to generalize new targets. This was done with the sequence-similarity-based splitting strategy, in which the pairwise sequence similarity of all proteins was computed by MMSeqs2  \cite{steinegger2017mmseqs2} and all PDBs up to a certain similarity threshold were grouped together. The similarity cutoff in this case was 35 \% and splitting was done until a 75:25 train test ratio was achieved. Ultimately, the train set contained 2766 proteins, the test set contained 1083 proteins, and the validation set contained 305 proteins.

For the MD simulation the co-crystallized small molecules, waters and ions were removed with Schrödinger’s Protein Preparation Wizard \cite{madhavi2013protein}. Simulations were performed with Schrödinger’s Desmond according to its standard NPT relaxation protocol \cite{bowers2006scalable}. The production simulations ran at 300K for 20 ns each, with a $50\
\text{kcal mol}^{-1} \, \text{\AA}^{-2}$ large positional restraint applied to all heavy atoms. Every 20 ps snapshots were taken, resulting in 1000 frames per simulation for analysis. Using VMD \cite{humphrey1996vmd} the MD simulations were aligned and recentered and the hydration sites calculated from each simulation trajectory using the WATsite program. The water molecules were tracked throughout the simulation and their density is discretized onto a grid with resolution of 0.25\AA. A hierarchical clustering algorithm clustered the grid into hydration sites with a radius of 1\AA. Water molecules belonging to each hydration site are extracted from the simulation allowing for the calculation of thermodynamic properties. The change in enthalpy ($\Delta$H)  was determined based on the average sum of van der Waals and electrostatic interactions between each water molecule inside a hydration site with the protein and other water molecules compared to the value in bulk solvent. Meanwhile, the entropy was estimated by integration of the external mode probability density function (PDF, $p_{ext}(q)$) of the water molecules’ translational and rotational motions during the MD simulation via the following formula:
\begin{equation}
\Delta S_{\mathrm{hs}}=R \ln \left(\frac{C^{\circ}}{8 \pi^2}\right)-R \int p_{\mathrm{ext}}(q) \ln p_{\mathrm{ext}}(q) d q
\end{equation}
With R being the gas constant, $\text{C}^\circ$ the concentration of pure water (1 molecule/29.9 $\text{\AA}^3$). 
From entropy and enthalpy, the Gibbs free energy of the desolvation of hydration sites could be calculated using the following formula:
$\Delta G= \Delta H-T \Delta S$.
It represents the temperature (T) dependent change of entropy and enthalpy of a water molecule inside a hydration site, when it is transferred from the bulk solvent into the hydration site of the protein cavity \cite{yang2018calculation}.

\section{Results and Discussion}

\subsection{Hydration site prediction}
In the following, we present the results of the model discussed in section \ref{sec:location_prediction} on the test set (similar results for the training set are given in section \ref{sec:results_train_set} of the Appendix). We evaluate the prediction performance in terms of the \textbf{G}round \textbf{T}ruth \textbf{R}ecovery \textbf{R}ate (\textbf{GTRR}) and \textbf{P}rediction \textbf{H}it \textbf{R}ate (\textbf{PHR}) as defined in Appendix \ref{sec:evaluation_metrics}. Table \ref{table:total_evaluation} shows the GTRR and PHR for a cutoff value of $0.5$ {\AA} and $1.0$ {\AA}. The model recovers $59\%$ of the true hydration site at a cutoff level of $0.5$ {\AA} divergence. If we increase the cutoff to $1.0$ {\AA}, then $80.2\%$ of hydration sites are recovered. On the other hand, a proportion of $48.3\%$ of predictions is closer than $0.5$ {\AA} to a true hydration site and $65.9\%$ are closer than $1.0$ {\AA}.

\begin{table}[h!]
    \centering
    \begin{tabular}{c|c c}
        \hline
        \textbf{Cutoff} & \textbf{Ground Truth Recovery Rate} & \textbf{Prediction Hit Rate} \\
        \hline
        0.5 & 59.0\% & 48.3\% \\
        1.0 & 80.2\% & 65.9\% \\
        \hline
    \end{tabular}
	\captionsetup{aboveskip=5pt, belowskip=5pt}
    \caption{Ground truth recovery rate and prediction hit rate on WATsite test set}
	\label{table:total_evaluation}
\end{table}

We consider a hydration site to be in the \textit{first layer} if it is in no further distant than $3.5$ {\AA} from a non-hydrogen atom of the protein. Otherwise the hydration site is in the second layer. Table \ref{tab:gtrr_by_layer} shows the GTRR for first and second layer hydration sites at cutoff levels of $0.5$ and $1.0$ {\AA}. We observe that the model very successfully predicts first layer hydration site coordinates with a GTRR of $62.5\%$ and $84.5\%$ for and $0.5$ and $1.0$ {\AA} respectively. The second layer hydration sites are substantially more challenging to predict correctly, as only $15.2\%$ and $26.9\%$ of true hydration sites are recovered at a tolerance cutoff of $0.5$ and $1.0$ {\AA} respectively.
These results indicate a clear distinction in the model's predictive accuracy between first and second shell hydration sites.
First layer hydration sites are strongly influenced by electrostatic and steric effects, often forming direct hydrogen bonds with the protein, making them more predictable.
Second layer hydration sites have reduced accuracy, like due to their weaker, indirect interactions with the proteins.
This leads to greater mobility of the water and less localized water density.

\begin{table}[h!]
    \centering
    \begin{tabular}{c|c c c}
        \hline
        \textbf{Cutoff} & \textbf{GTRR first layer}& \textbf{GTRR second layer} \\
        \hline
        0.5 & 62.5\% & 15.2\%\\
        1.0 & 84.5\% & 26.9\%\\
        \hline
    \end{tabular}
	\captionsetup{aboveskip=5pt, belowskip=5pt}
    \caption{Ground truth recovery rate (GTRR) for first and second layer}
    \label{tab:gtrr_by_layer}
\end{table}

We further investigate the recovery performance for hydration sites differentiated based on occupancy in Table \ref{tab:gtrr_total}. As one would naturally expect, the GTRR increases with higher occupancy, meaning more stable hydration sites are easier to predict. The GTRR is highest for hydration sites of an occupancy between $0.8$ and $0.9$ with $70.0\%$ and $91.0\%$ for a divergence cutoff of $0.5$ and $1.0$ {\AA} respectively. For the hydration sites of highest occupancy (between $0.9$ and $1.0$) the model receives almost the same performance with $69.7\%$ and $90.8\%$ respectively.

\begin{table}[h!]
    \centering
    \begin{tabular}{c|c c c c c}
        \hline 
        \multicolumn{1}{c|}{} & \multicolumn{5}{c}{\textbf{Occupancy}} \\
        \multicolumn{1}{c|}{} & \textbf{[0.5,0.6]} & \textbf{[0.6,0.7]} & \textbf{[0.7,0.8]} & \textbf{[0.8,0.9]} & \textbf{[0.9,1.0]} \\
        \hline 
        \textbf{Cutoff} & & & & & \\
        $r = 0.5$ & 42.1\% & 57.2\% & 65.1\% & 70.0\% & 69.7\% \\
        $r = 1.0$ & 62.4\% & 79.4\% & 86.9\% & 91.0\% & 90.8\% \\
        $r = 1.5$ & 70.4\% & 86.3\% & 92.5\% & 95.7\% & 94.4\% \\
        $r = 2.0$ & 75.1\% & 89.5\% & 94.8\% & 97.3\% & 95.6\% \\
        \hline 
    \end{tabular}
    \captionsetup{aboveskip=5pt, belowskip=5pt}
    \caption{Ground truth recovery rate at different occupancy levels}
    \label{tab:gtrr_total}
\end{table}

As shown in Table \ref{tab:gtrr_layer1_by_occupancy}, the results signficantly improve for low occupancy rates, if we just consider the hydration sites of the first layer.

\begin{table}[h!]
    \centering
    \begin{tabular}{c|c c c c c}
        \hline 
        \multicolumn{1}{c|}{} & \multicolumn{5}{c}{\textbf{Occupancy}} \\
        \multicolumn{1}{c|}{} & \textbf{[0.5,0.6]} & \textbf{[0.6,0.7]} & \textbf{[0.7,0.8]} & \textbf{[0.8,0.9]} & \textbf{[0.9,1.0]} \\
        \hline 
        \textbf{Cutoff} & & & & & \\
        $r = 0.5$ & 48.5\% & 60.6\% & 67.0\% & 70.9\% & 69.9\% \\
        $r = 1.0$ & 71.0\% & 83.6\% & 89.0\% & 91.9\% & 90.9\% \\
        $r = 1.5$ & 79.4\% & 90.5\% & 94.5\% & 96.5\% & 94.5\% \\
        $r = 2.0$ & 83.7\% & 93.3\% & 96.5\% & 97.9\% & 95.7\% \\
        \hline 
    \end{tabular}
    \captionsetup{aboveskip=5pt, belowskip=5pt}
    \caption{First layer waters: Ground truth recovery rate at different occupancy levels}
    \label{tab:gtrr_layer1_by_occupancy}
\end{table}

\subsection{Thermodynamic Profiling}

In the following, we evaluate our model for thermodynamic profile prediction as introduced in section \ref{sec:enthalpy_entropy_prediction}. Given the coordinates of the true hydration sites, we investigate how accurately the model can recover the enthalpy and entropy values given by the explicit-water MD simulations with WATsite. A case study where we predict thermodynamic properties based on predicted hydration site locations is provided in subsection \ref{sec:mup}.

Table \ref{tab:cor_mse_test} shows the results for both enthalpy and entropy in terms of mean squared error and Pearson correlation on the test set. The correlation for enthalpy and entropy is visualized in hexbin density plots in Figure \ref{fig:enthalpy_correlation} and \ref{fig:entropy_correlation} respectively.
We observe that given the true hydration site coordinates, the correlation between model predictions and the corresponding simulated target values is high, with $0.864$ for entropy and $0.839$ for enthalpy.
Interestingly, there seems to be an increasing uncertainty for predictions in the case of higher entropy values.
High-entropy waters are often characterized by greater mobility, weaker structural organization and fewer hydrogen bonds.
This leads to a broader distribution of possible locations and orientations of the water molecule.
Consequently, these less ordered hydration sites are harder to predict accurately.

\begin{table}[h!]
    \centering
    \begin{tabular}{c|c c c}
        & \textbf{Entropy}& \textbf{Enthalpy} \\
        \hline
	\textbf{MSE} & 0.218 & 0.38\\
	\textbf{Correlation} & 0.864& 0.839\\
    \end{tabular}
	\captionsetup{aboveskip=5pt, belowskip=5pt}
	\caption{Correlation and Means Squared Error (MSE) between prediction and ground truth for both entropy and enthalpy on the test set}
	\label{tab:cor_mse_test}
\end{table}



\begin{figure}[h]
    \centering
    \includegraphics[width=\linewidth]{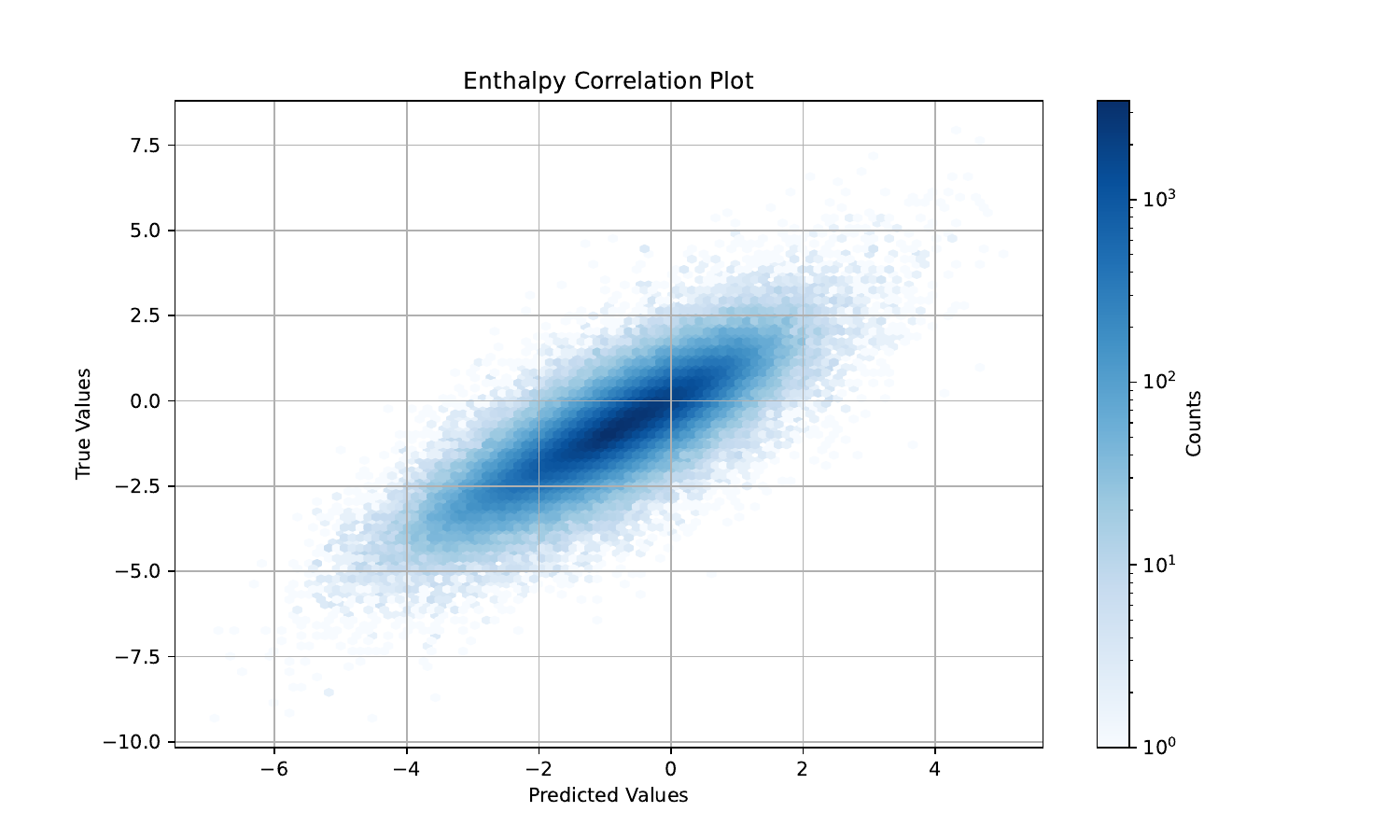}
	\captionsetup{aboveskip=5pt, belowskip=5pt}
    \caption{Hexbin plot showing the correlation between the enthalpy predictions and the enthalpy values obtained from WATsite for the test set.}
    \label{fig:enthalpy_correlation}
\end{figure}

\begin{figure}[h]
    \centering
    \includegraphics[width=\linewidth]{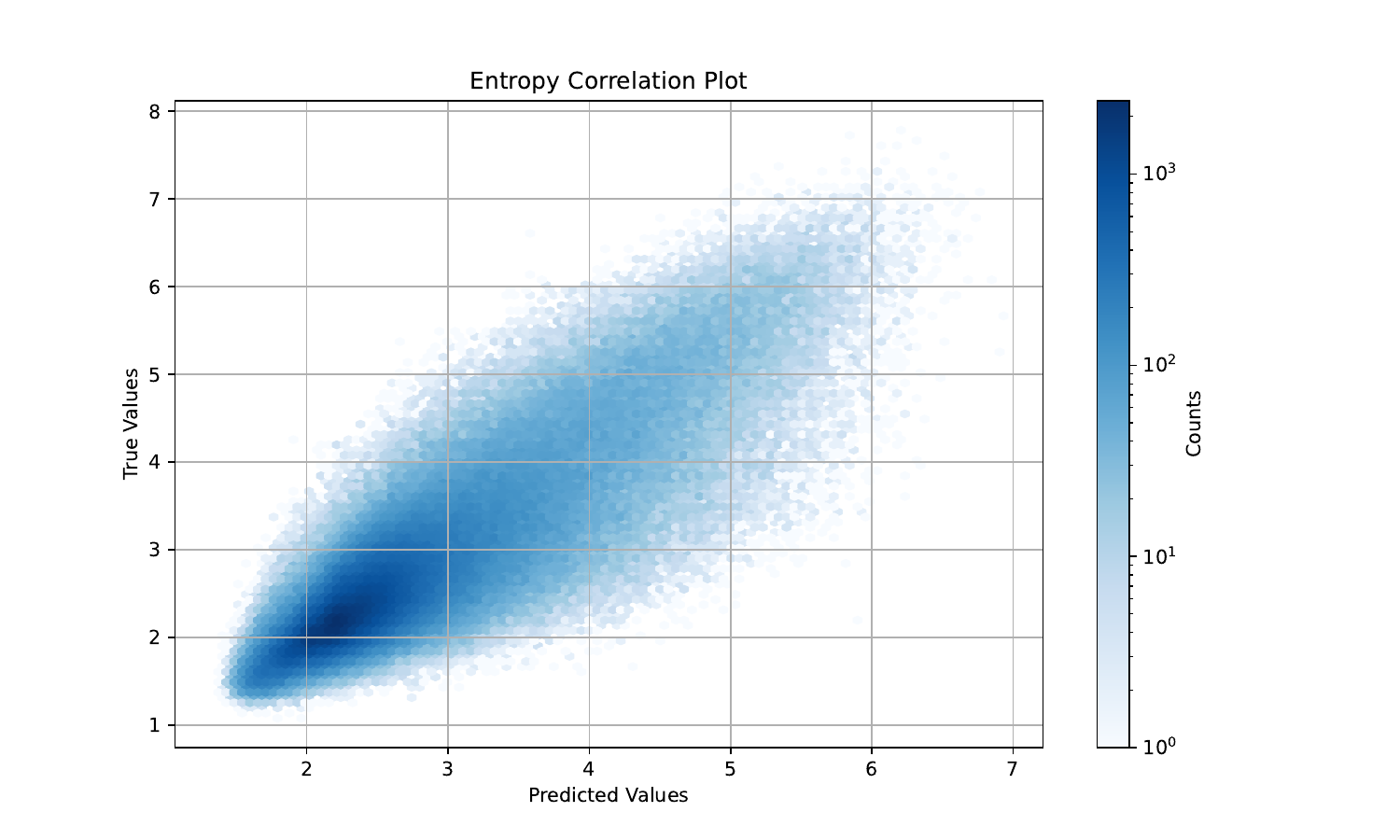}
    \caption{Hexbin plot showing the correlation between the entropy predictions and the entropy values obtained from WATsite for the test set.}
    \label{fig:entropy_correlation}
\end{figure}

\subsection{Case Studies}
In order to investigate the accuracy and behavior of the model more intently, we looked at five specific systems: Disulphide Catalyst DsbA, Triosephosphate Isomerase, Heat Shock Protein 90 (HSP90), Clarin-2, and Major Urinary Protein (MUP).
The first two focus on conserved water networks important for structure and function as well as the effect of point mutations.
The HSP90 example focuses on conformational changes ranging from side-chain rotation to loop reorganization.
Clarin-2 was used to test the ability of the model to predict hydration information using AlphaFold3 structures of unresolved membrane-bound proteins.
Finally, in the MUP case study we apply thermodynamic predictions to the case of protein-ligand binding and show that a strong correlation can be established between the estimated free energy of desolvation and experimentally measured affinity values for a series of ligands.

\subsubsection{Disulphide Catalyst DsbA}

\begin{figure}[h]
    \centering
    \includegraphics[width=\linewidth]{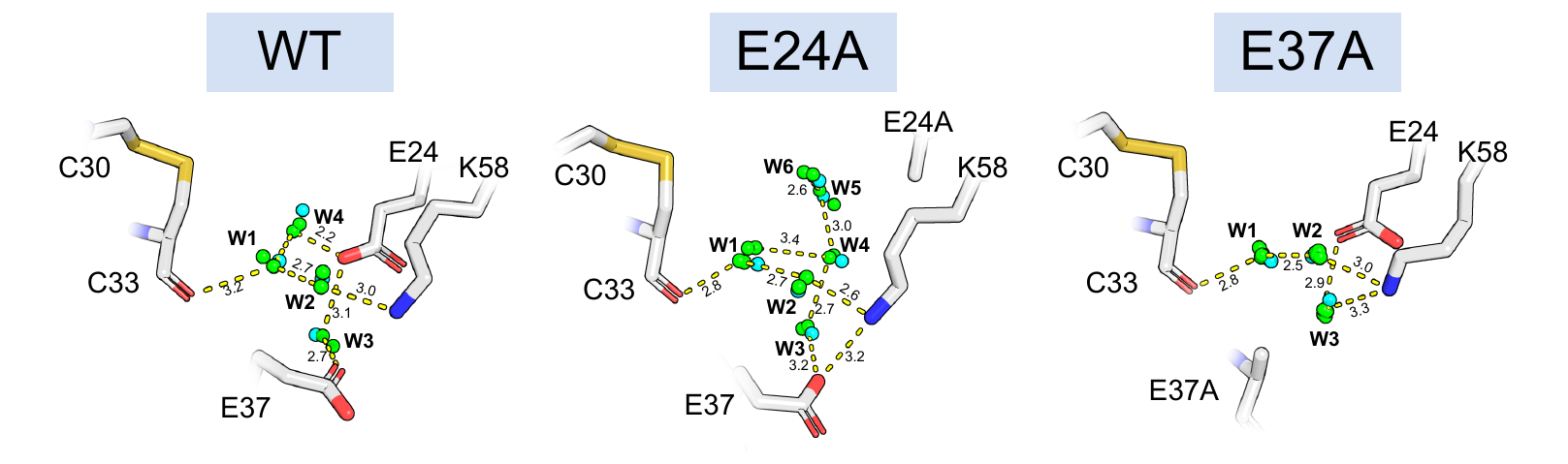}
    \caption{Buried water network important for the activity of disulphide catalyst DsbA. We evaluated the wildtype (PDB: 5QO9) as well as two mutants known to disrupt the conserved water network, E24A (PDB: 8EQR) and E37A (PDB: 8EQQ). Crystal waters are shown from all superimposed chains in green. Hydration site predictions are shown in cyan. Relevant distances are shown and labelled with their length in Angstrom. Predictions beyond the selected crystal hydration sites are not shown.}
    \label{fig:dsba}
\end{figure}

Disulphide Catalyst DsbA is an important enzyme for the formation of disulphide bonds, which are present in many proteins entering the secretary pathway \cite{robinson2020mechanisms}.
Importantly, disulphide catalysts contain a conserved buried water network within the active site that is essential for it's function.
It is believed that the water network acts as a "proton wire" that shuffles protons from the bulk solvent to the catalytic cysteine residues.
Previous studies have investigated the effects of point mutations within the active site using x-ray crystallography \cite{wang2023buried}.
In this experiment, we aimed to test if our model could correctly generate the conserved water network structure and respond accurately to point mutations.
The results, shown in Figure \ref{fig:dsba}, show a high level of accuracy for both the wild type (WT) and both point mutations (E24A and E37A).
In each case, we were able to correctly identify all of the hydration sites within 1.0 {\AA}.
These results demonstrate the model is capable of predicting structurally and functionally important water networks with a high degree of accuracy.
The model also appears robust and generalizable by it's ability to accurately respond to point mutations.
A common fault of deep-learning-based models is their tendency to overfit and inability to behave accurately outside of the training data distribution \cite{liu2021towards}.
However, the results indicate our model is not overfit on any particular protein folds and can respond to small changes in protein structure.
Lastly, the predicted hydration sites appear to be physically valid with no steric clashes and reasonable interatomic distances.

\subsubsection{Triosephosphate Isomerase}

\begin{figure}
    \centering
    \includegraphics[width=\linewidth]{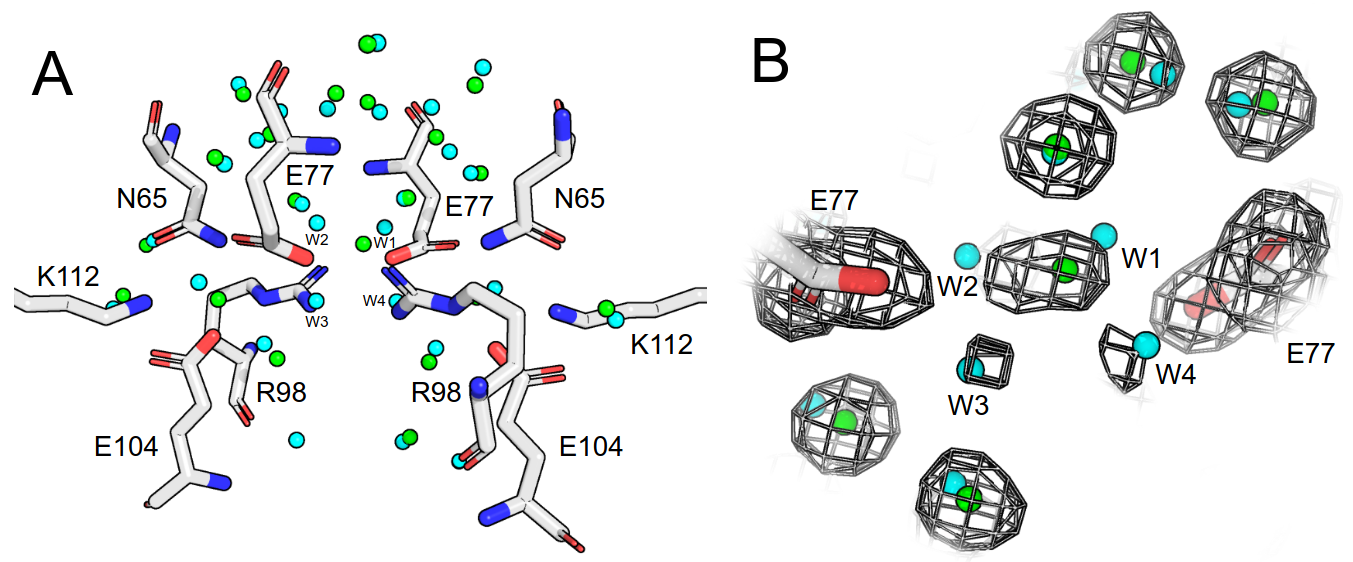}
    \caption{Conserved water network at the dimer interface of triosephosphate isomerase. A) Residues at interface of subunits A (right) and B (left) with crystal waters (green) and predicted hydration sites (cyan). B) Focused view of the single failure at center of interface with 2fo-fc map shown as mesh with isovalue of 1.0.}
    \label{fig:2jk2}
\end{figure}

Triosephosphate isomerase is a dimeric enzyme involved in the glycolysis pathway and is found in nearly all organisms \cite{triosephosphate2014}.
Like in the previous case study, triosephosphate isomerase also contains a conserved water network.
However, in this case the water network is present at the interface of the homodimer structure and not buried within the protein.
The predictions showed very strong agreement with experimental data with 20/21 crystal waters predicted within 1.0 {\AA} (Figure \ref{fig:2jk2}A).
Upon further inspection of the single failure, our proposed hydration sites occupy unmodelled regions of electron density, potentially offering a better fit than the PDB deposited structure (Figure \ref{fig:2jk2}B).

\subsubsection{Heat Shock Protein 90}

\begin{figure}[h]
    \centering
    \includegraphics[width=\linewidth]{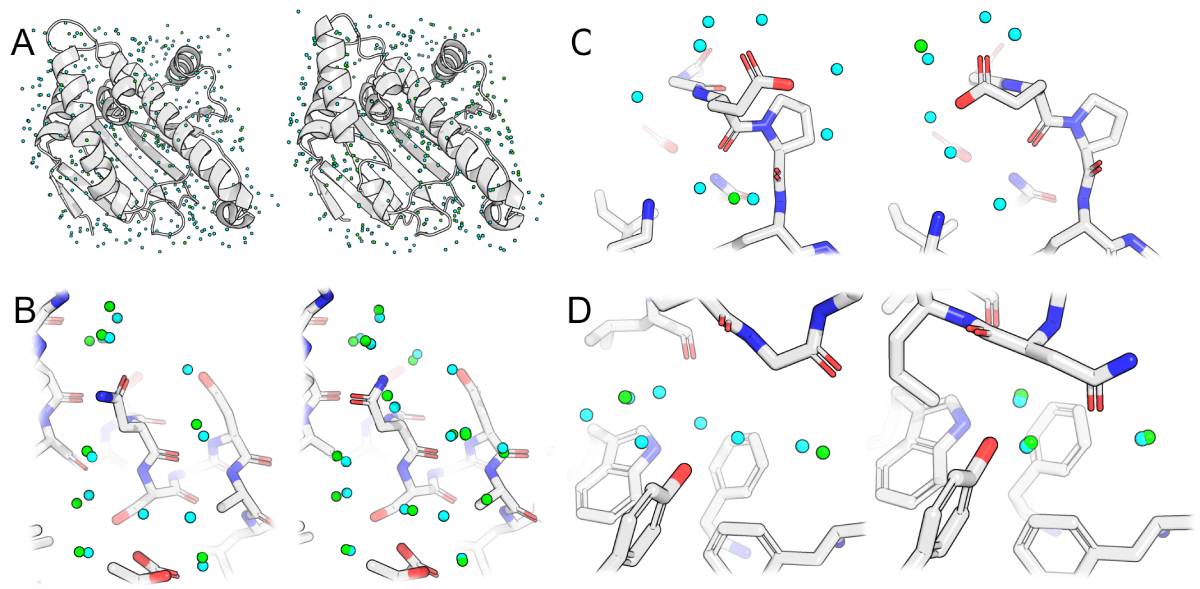}
    \caption{Comparison of HSP90 in two conformations from PDBs 2QFO and 2WI7 on the left and right respectively. A) Full protein structures with all crystal waters (green) and predictions (cyan). B) Focused view on region with similar conformation in both structures and strong agreement with crystal waters. C) Example of flexible side-chain in different conformations with corresponding predicted waters. D) Several hydration sites occupying the open pocket of 2QFO are displaced or stabilized by the $\alpha$-helix3 side-chains seen in the 2WI7 loop-in structure.}
    \label{fig:hsp90}
\end{figure}

The location of hydration sites is highly dependent on protein conformation \cite{yang2014analysis, yang2016dissecting}.
Therefore, it is essential that the model is capable of responding to changes in the protein ranging from small perturbations to large-scale conformational changes.
To this end, we predicted hydration sites for HSP90 in two conformations which exhibits both small and large conformational change (Figure \ref{fig:hsp90}).
In regions where the conformation is similar, the model predicted similar water networks that exhibited strong agreement with crystal water locations (Figure \ref{fig:hsp90}B).
This confirms that distant changes in protein structure does not result in undue changes to the hydration site prediction.
In the case of a flexible side-chain that can adopt multiple possible conformations, the model was able to adapt to the change, maintaining proper hydrogen bond distances and avoiding steric clashes (Figure \ref{fig:hsp90}C).
This finding demonstrates the robustness of our model in responding to small changes and accurately modelling atomic interactions.
This also shows that our model is not overfit to any particular conformation and the network hasn't simply memorized the placement of hydration sites based on sequence alone.
HSP90 also exhibits a larger conformational change among it's $\alpha$-helix3.
In 2QFO the $\alpha$-helix is intact while in 2WI7 the helix unfolds in the middle and adopts a "loop-in" conformation.
The model accurately responds to the large conformational change and is able to detect both the displacement and stabilization of several sites, further supported by crystallographic waters (Figure \ref{fig:hsp90}D).

\subsubsection{Clarin-2}

\begin{figure}
    \centering
    \includegraphics[width=\linewidth]{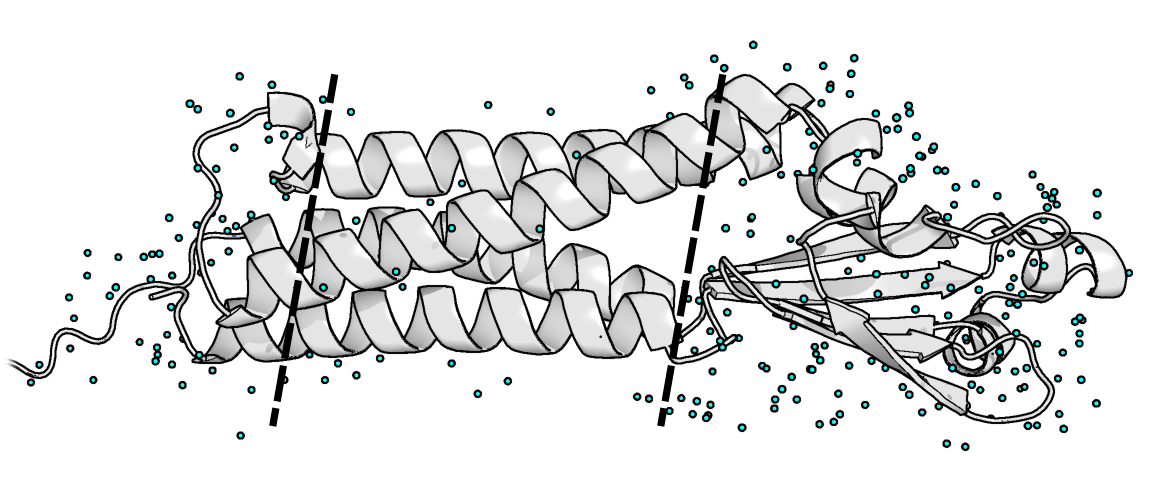}
    \caption{Hydration site predictions for membrane-bound protein Clarin-2 generated via AlphaFold3. The predictions show that they are robust to different molecular environments and can accurately predict the number of waters depending on the environment. In the hydrophobic, membrane-bound core of the protein, there are very few predicted waters. However, the hydrophilic intra- and extra-cellular sides had a large number of predicted sites.}
    \label{fig:clrn}
\end{figure}

One exciting new opportunity to test our model is on the host of computationally predicted structures which have not been experimentally determined.
This is especially interesting since leading protein structure methods, such as AlphaFold3 and RosettaFold-AllAtom, do not consider water molecules in their predictions \cite{abramson2024accurate, krishna2024generalized}.
Thus, there exists thousands of protein structures for which important water locations cannot be predicted easily.
To demonstrate this advantage, we used AlphaFold3 to predict the structure of a membrane-bound protein Clarin-2, which has no structure contained in the PDB.
The structure of membrane-bound proteins are often difficult in general to predict due to their inability to crystallize, although recent advances have enabled the determination of some membrane-bound proteins \cite{carpenter2008overcoming, kermani2021guide}.
The Clarin-2 protein is essential for hearing function by maintaining stereocilia integrity and could have therapeutic applications \cite{dunbar2019clarin, mendia2024clarin}.
The AlphaFold3 structure seems reasonable according to the confidence metrics and visual inspection; the structure contains the expected four transmembrane domains as well as interfaces on the intra and extra-cellular sides \cite{adato2002ush3a}.
An interesting finding was observed after predicting the water locations, shown in Figure 
\ref{fig:clrn}.
A clear distinction in the number of predicted waters can be seen between the exposed intra/extra-cellular regions and the membrane-bound region.
This result demonstrates the ability of our model to not only accurately predict the locations of waters surrounding a protein but to also accurately predict their likelihood.
Furthermore, it shows that the model is able to capture essential aspects of physics such as hydrophobicity.

\subsubsection{Major Urinary Protein (MUP)}
\label{sec:mup}

The free energy of desolvation is a major driving force for protein-ligand binding.
In order to test the performance of our thermodynamic profiling in the context of ligand binding, 12 cocrystallized ligands to MUP were overlaid with the predicted hydration sites and the free energy of desolvation was calculated based on the displaced water molecules (Table \ref{tab:mup})\footnotemark[1]

\begin{figure}[h!]
    \centering
    \includegraphics[width=\linewidth]{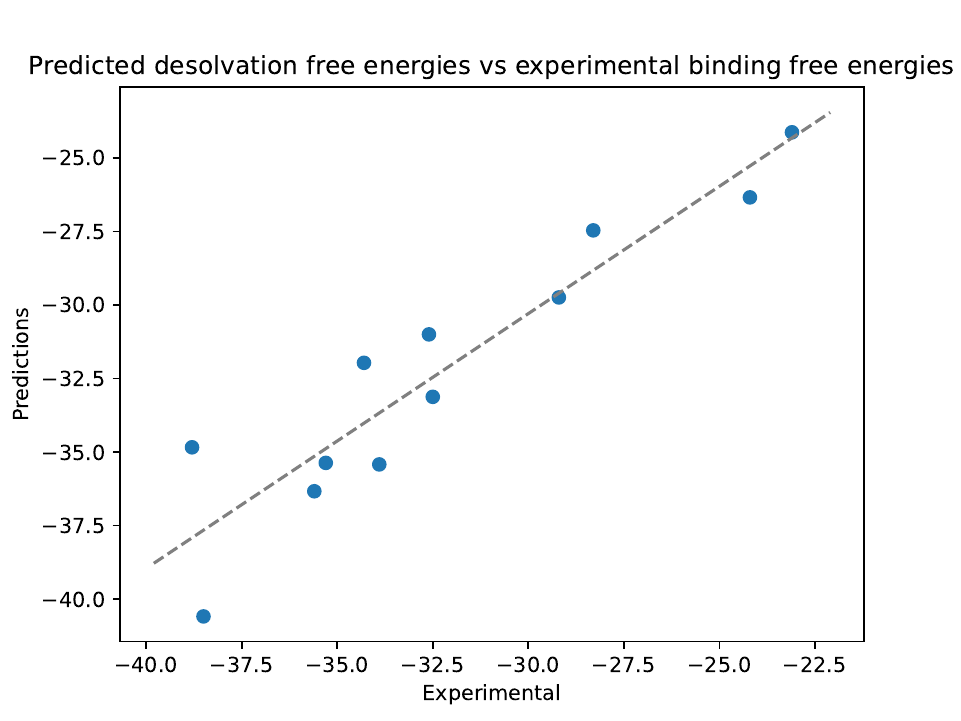}
    \caption[A]{Regression plot comparing predicted desolvation free energy and binding free energy.}
    \captionsetup{aboveskip=5pt, belowskip=5pt}
    \label{fig:displacement_plot}
\end{figure}

While the desolvation free energy is only one component of ligand binding, these ligands are largely hydrophobic and share a single carboxylic acid moiety, and therefore the predominant difference between them is the number and position of displaced water molecules.
Figure \ref{fig:displacement_plot} shows a regression plot between the predicted energies \footnotemark[1] for all ligands in the data set and the experimentally measured values. At a water displacement tolerance level of $2.4$ {\AA}, we obtain a Pearson R correlation of $0.930$.

\footnotetext[1]{Shown are the transformed predictions obtained by a linear regression of the experimental values on the model predictions.}
This significant correlation is superior to MM-GB/SA results we obtained for the same complexes, where almost no correlation was found (Pearson R < 0.05).

\begin{table}[]
    \centering
    \resizebox{\textwidth}{!}{
        \begin{tabular}{|c|c|c|c|c|c|c|c|c|c|c|c|c|}
        \hline PDB ID & 1ZND & 1ZNE & 1ZNG & 1ZNH & 1ZNK & 1QY1 & 1QY2 & 1IO6 & 1I06 & 1I06 & 1IO6 & 1I06 \\
        \hline Ligand & PE9 & HE2 & HE4 & OC9 & F09 & IBMP & IPMP & SBT & PT & IPT & ET & MT \\
	\hline Method & -24.1 & -27.5 & -33.1 & -36.3 & -34.8 & -40.5 & -35.4 & -35.4 & -31.9 & -31.0 & -29.7 & -26.3 \\
        WATsite & -18.1 & -22.0 & -17.9 & -31.1 & -31.1 & -31.5 & -35.8 & -30.6 & -34.1 & -30.6 & -24.2 & -16.3 \\
        Experimental & -23.1 & -28.3 & -32.5 & -35.6 & -38.8 & -38.5 & -33.9 & -35.3 & -34.3 & -32.6 & -29.2 & -24.2 \\
        MM-GB/SA & -28.5 & -33.6 & -21.8 & -17.8 & -32.7 & -24.8 & -22.8 & -33.0 & -29.3 & -28.8 & -25.1 & -23.5 \\
        \hline
        \end{tabular}
    }
    \caption{Binding free energies from hydration site displacement of 12 MUP ligands. Free energies are reported in kJ/mol. For ligands following SBT, structures were obtained by removing specified carbon atoms from the crystal structure of SBT.}
    \label{tab:mup}
\end{table}


\section{Conclusion}
We introduced a new method for hydration site prediction and thermodynamic profiling based on an equivariant deep neural network trained on a large corpus of MD data.
Our results demonstrate that our model is fast, accurate, and robust to chemical and biological perturbations.
Other hydration site profiling tools can take several hours for just a single protein conformation due to the expensive explicit-water simulations.
However, by leveraging our model we can make predictions on the seconds time scale, enabling high-throughput hydration site prediction and profiling.
This is a crucial advancement to predict hydration sites for flexible proteins in different conformations or for the integration of hydration site prediction into deep learning models for molecular structure prediction.
Our results further demonstrated that the model is highly accurate and can reproduce results from molecular dynamics and experimentally resolved structures with high fidelity.
Furthermore, thermodynamic predictions from our model showed strong agreement with those obtained from MD and experimentally measured affinities for a series of ligands.
We investigated the robustness and generalization of our model since this is a common failure mode seen in deep learning models.
Specifically, we evaluated the model under different perturbations to the protein input such as conformational changes and single-point mutations, then confirmed our predictions against the known water locations.
The model showed no signs of overfitting and was capable of responding accurately to small changes in the overall protein structure. 
In summary, our model offers several key advantages over existing hydration site prediction tools.
The ability to quickly and accurately predict hydration site locations and thermodynamics has a range of potential applications, from small molecule drug design and lead optimization to peptide and protein design.
Our model could also be integrated with other emerging technologies such as deep learning models for co-folding, dynamics, and free energy estimation.

\subsection*{Acknowledgment}
The work was partially supported by the Swiss National Science Foundation (Project number: 310030\_197629) and the Novartis Research Foundation. 

\subsection*{Data availability}
\begin{itemize}
	\item{ The code for this work is provided here: \href{https://github.com/lillgroup/HydrationSitePrediction}{https://github.com/lillgroup/HydrationSitePrediction}}
	\item{ The data is available here: \href{https://zenodo.org/records/14182834}{https://zenodo.org/records/14182834}}
\end{itemize}

\clearpage

\bibliography{manuscript}

\clearpage
\appendix

\section{Graphics}

\begin{figure}[h!]
    \centering
    \includegraphics[width=0.9\linewidth]{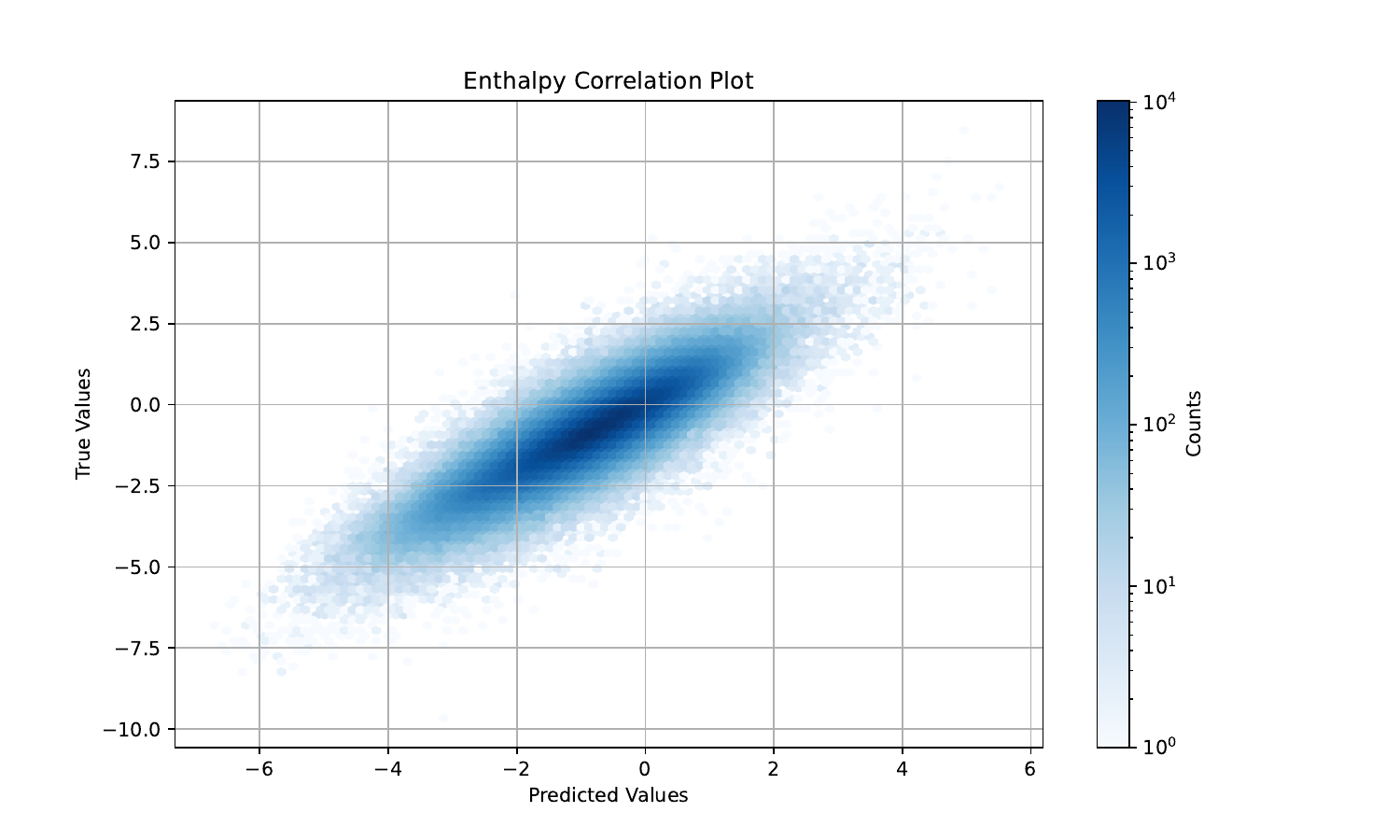}
	\captionsetup{aboveskip=5pt, belowskip=5pt}
    \caption{Hexbin plot showing the correlation between the enthalpy predictions and the enthalpy values obtained from WATsite for the train set.}
    \label{fig:enthalpy_correlation}
\end{figure}

\begin{figure}[h!]
    \centering
    \includegraphics[width=0.9\linewidth]{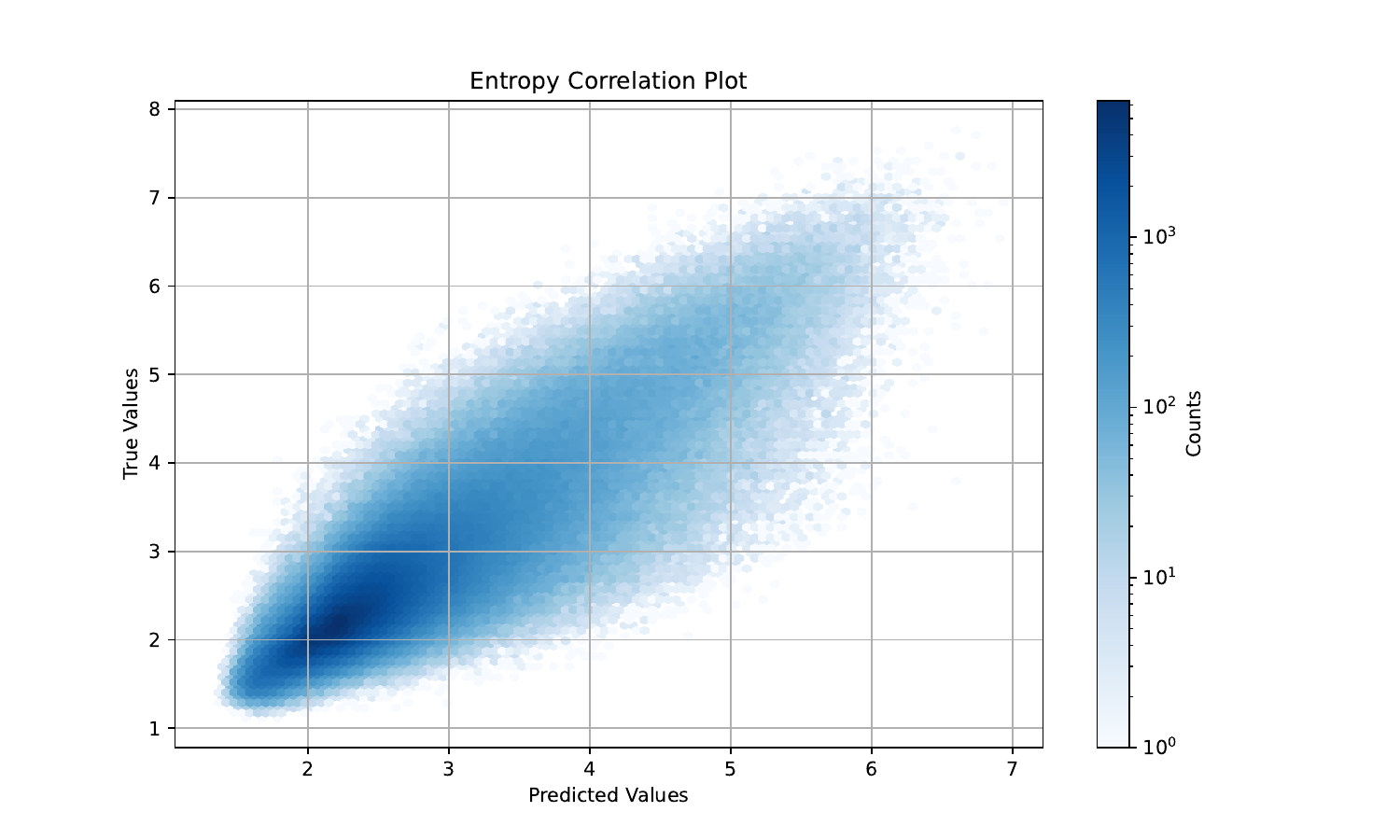}
    \caption{Hexbin plot showing the correlation between the entropy predictions and the entropy values obtained from WATsite for the train set.}
    \label{fig:entropy_correlation}
\end{figure}
\clearpage

\section{Tables}

\subsection{Further results on test set}

\begin{table}[h!]
    \centering
    \begin{tabular}{c|c c}
        \hline
        \textbf{Cutoff} & \textbf{Ground Truth Recovery Rate} & \textbf{Prediction Hit Rate} \\
        \hline
	0.5 & 37.0\% & 6.26\% \\
	1.0 & 68.3\% & 11.5\% \\
        \hline
    \end{tabular}
	\captionsetup{aboveskip=5pt, belowskip=5pt}
    \caption{Ground truth recovery rate and prediction hit rate on xray data}
\end{table}

\begin{table}[h!]
    \centering
    \label{tab:gtrr_second_layer}
    \begin{tabular}{c|c c c c c}
        \hline 
        \multicolumn{1}{c|}{} & \multicolumn{5}{c}{\textbf{Occupancy}} \\
        \multicolumn{1}{c|}{} & \textbf{[0.5,0.6]} & \textbf{[0.6,0.7]} & \textbf{[0.7,0.8]} & \textbf{[0.8,0.9]} & \textbf{[0.9,1.0]} \\
        \hline 
        \textbf{Cutoff} & & & & & \\
        $r = 0.5$ & 11.4\% & 17.6\% & 22.8\% & 29.4\% & 38.5\% \\
        $r = 1.0$ & 20.8\% & 30.6\% & 39.8\% & 50.7\% & 63.5\% \\
        $r = 1.5$ & 26.6\% & 38.0\% & 48.7\% & 59.4\% & 74.5\% \\
        $r = 2.0$ & 33.3\% & 45.6\% & 56.2\% & 68.2\% & 82.2\% \\
        \hline 
    \end{tabular}
    \captionsetup{aboveskip=5pt, belowskip=5pt}
    \caption{Second layer waters: Ground truth recovery rate at different occupancy levels}
\end{table}

\subsection{Results on train set}
\label{sec:results_train_set}
For reference we provide the results for the training set as well.

\begin{table}[h!]
    \centering
    \begin{tabular}{c|c c}
        \hline
        \textbf{Cutoff} & \textbf{Ground Truth Recovery Rate} & \textbf{Prediction Hit Rate} \\
        \hline
        0.5 & 62.7\% & 50.2\% \\
        1.0 & 82.1\% & 66.0\% \\
        \hline
    \end{tabular}
	\captionsetup{aboveskip=5pt, belowskip=5pt}
    \caption{Ground truth recovery rate and prediction hit rate}
\end{table}

\begin{table}[h!]
    \centering
    \begin{tabular}{c|c c c}
        \hline
        \textbf{Cutoff} & \textbf{GTRR first layer}& \textbf{GTRR second layer} \\
        \hline
        0.5 & 66.2\% & 16.7\%\\
        1.0 & 86.3\% & 28.2\%\\
        \hline
    \end{tabular}
	\captionsetup{aboveskip=5pt, belowskip=5pt}
    \caption{Ground truth recovery rate (GTRR) for first and second layer}
\end{table}

\begin{table}[h!]
    \centering
    \begin{tabular}{c|c c c c c}
        \hline 
        \multicolumn{1}{c|}{} & \multicolumn{5}{c}{\textbf{Occupancy}} \\
        \multicolumn{1}{c|}{} & \textbf{[0.5,0.6]} & \textbf{[0.6,0.7]} & \textbf{[0.7,0.8]} & \textbf{[0.8,0.9]} & \textbf{[0.9,1.0]} \\
        \hline 
        \textbf{Cutoff} & & & & & \\
        $r = 0.5$ & 45.3\% & 61.1\% & 69.2\% & 74.6\% & 75.4\% \\
        $r = 1.0$ & 65.1\% & 81.8\% & 89.1\% & 93.0\% & 93.3\% \\
        $r = 1.5$ & 72.2\% & 87.5\% & 93.6\% & 96.6\% & 95.9\% \\
        $r = 2.0$ & 76.2\% & 90.1\% & 95.4\% & 97.8\% & 96.7\% \\
        \hline 
    \end{tabular}
    \captionsetup{aboveskip=5pt, belowskip=5pt}
    \caption{Ground truth recovery rate at different occupancy levels}
\end{table}

\begin{table}[h!]
    \centering
    \begin{tabular}{c|c c c c c}
        \hline 
        \multicolumn{1}{c|}{} & \multicolumn{5}{c}{\textbf{Occupancy}} \\
        \multicolumn{1}{c|}{} & \textbf{[0.5,0.6]} & \textbf{[0.6,0.7]} & \textbf{[0.7,0.8]} & \textbf{[0.8,0.9]} & \textbf{[0.9,1.0]} \\
        \hline 
        \textbf{Cutoff} & & & & & \\
        $r = 0.5$ & 51.7\% & 64.4\% & 70.9\% & 75.3\% & 75.6\% \\
        $r = 1.0$ & 73.5\% & 85.8\% & 90.9\% & 93.8\% & 93.4\% \\
        $r = 1.5$ & 80.9\% & 91.5\% & 95.3\% & 97.3\% & 96.0\% \\
        $r = 2.0$ & 84.5\% & 93.7\% & 96.9\% & 98.3\% & 96.8\% \\
        \hline 
    \end{tabular}
    \captionsetup{aboveskip=5pt, belowskip=5pt}
    \caption{First layer waters: Ground truth recovery rate at different occupancy levels}
\end{table}

\begin{table}[h!]
    \centering
    \begin{tabular}{c|c c c c c}
        \hline 
        \multicolumn{1}{c|}{} & \multicolumn{5}{c}{\textbf{Occupancy}} \\
        \multicolumn{1}{c|}{} & \textbf{[0.5,0.6]} & \textbf{[0.6,0.7]} & \textbf{[0.7,0.8]} & \textbf{[0.8,0.9]} & \textbf{[0.9,1.0]} \\
        \hline 
        \textbf{Cutoff} & & & & & \\
        $r = 0.5$ & 12.3\% & 19.8\% & 27.8\% & 35.5\% & 39.6\% \\
        $r = 1.0$ & 21.6\% & 32.8\% & 44.8\% & 54.5\% & 64.5\% \\
        $r = 1.5$ & 27.3\% & 39.1\% & 52.2\% & 63.1\% & 72.8\% \\
        $r = 1.5$ & 33.4\% & 46.1\% & 58.9\% & 69.5\% & 79.2\% \\
        \hline 
    \end{tabular}
	\captionsetup{aboveskip=5pt, belowskip=5pt}
\caption{Second layer waters: Ground truth recovery rate at different occupancy levels}
\end{table}

\begin{table}[h!]
    \centering
    \begin{tabular}{c|c c c}
        & \textbf{Entropy}& \textbf{Enthalpy} \\
        \hline
	\textbf{MSE} & 0.186 & 0.279\\
	\textbf{Correlation} & 0.867& 0.870\\
    \end{tabular}
	\captionsetup{aboveskip=5pt, belowskip=5pt}
	\caption{Correlation and Means Squared Error (MSE) between prediction and ground truth for both entropy and enthalpy on the training set}
	\label{tab:cor_mse_train}
\end{table}

\section{Evaluation measures}
\label{sec:evaluation_metrics}
In order to evaluate the quality of the hydration site coordinate predictions of our model we need to define two evaluation measure:  One measuring how many true hydration sites were discovered and a second one setting this into context by measuring how many predictions were reasonable.

Let $m,n\in \nn$ and $(a_{i})_{i=1}^{m}\in \rr^{3}$ be the “ground truth” points and $(b_{j})_{j=1}^{n}\in \rr^{3}$ be the “predictions“. Let $r\in \rr_{+}$ be the ”cutoff radius“. Let us agree to the following definitions

\begin{eqnarray}
	L^{r}_{GTRR}&:=& \frac{1}{m}\sum_{i=1}^{m} \left( 1-\prod_{j=1}^{n}\mathds{1}_{]r,\infty [}\left( ||a_{i}-b_{j}||_{2} \right) \right)	\\
	L^{r}_{PHR}&:=& \frac{1}{n} \sum_{j=1}^{n} \left( 1-\prod_{i=1}^{m}\mathds{1}_{]r,\infty [}\left( ||a_{i}-b_{j}||_{2} \right) \right)
\end{eqnarray}

We refer to $L^{r}_{GRR}$ as the \textbf{G}round \textbf{T}ruth \textbf{R}ecovery \textbf{R}ate (\textbf{GTRR}) and $L^{r}_{PHR}$ as the \textbf{P}rediction \textbf{H}it \textbf{R}ate (\textbf{PHR}).

In words, the \textit{ground truth recovery rate} is the proportion of ground truth elements that have at least one prediction within a radius $r$.
The \textit{prediction hit rate} is the proportion of predictions that are within a radius $r$ of \textit{any} ground truth element.

\end{document}